\documentclass[12pt, draftclsnofoot, onecolumn]{IEEEtran}

\usepackage{cite}
\usepackage{amsmath,amssymb,amsfonts}
\usepackage{algorithmic}
\usepackage{graphicx}
\usepackage{textcomp}
\usepackage{bm}
\usepackage{balance}
\usepackage{booktabs}
\usepackage{enumitem}
\usepackage{float}
\usepackage{url}

\newenvironment{linenomath}{}{}

\makeatletter
\AtBeginDocument{\DeclareMathVersion{bold}
\SetSymbolFont{operators}{bold}{T1}{times}{b}{n}
\SetMathAlphabet{\mathrm}{bold}{T1}{times}{b}{n}
\SetMathAlphabet{\mathit}{bold}{T1}{times}{b}{it}
\SetMathAlphabet{\mathbf}{bold}{T1}{times}{b}{n}
\SetMathAlphabet{\mathtt}{bold}{OT1}{pcr}{b}{n}
\SetSymbolFont{symbols}{bold}{OMS}{cmsy}{b}{n}
\renewcommand\boldmath{\@nomath\boldmath\mathversion{bold}}}
\makeatother

\def\BibTeX{{\rm B\kern-.05em{\sc i\kern-.025em b}\kern-.08em
    T\kern-.1667em\lower.7ex\hbox{E}\kern-.125emX}}

\makeatletter
\@ifundefined{linenomath}{\newenvironment{linenomath}{}{}}{}
\makeatother

\begin{document}

\title{Measurement-Driven Learning-Based Beam Selection for Hybrid Beamforming at 26.5 GHz}

\author{Kristian Drizari,
Konstantinos Maliatsos,
Vasileios Tsoulos,
Lefteris Tsipis,
Harris K. Armeniakos, and
Athanasios G. Kanatas%
\thanks{K. Drizari, V. Tsoulos,  H. K. Armeniakos and A. G. Kanatas are with the Department of Digital Systems, University of Piraeus, 18534 Piraeus, Greece (e-mail:
\{cdrizari, vtsoulos, harmen, kanatas\}@unipi.gr).}
\thanks{K. Maliatsos and L. Tsipis are with  the University of the Aegean, Samos, Greece (e-mail:
\{kmaliat,ltsipis\}@aegean.gr).}
\thanks{This work was supported in part by the University of Piraeus Research Centre, the EU SNS JU HORIZON Programme through iSEE-6G under Grant 101139291 and within the framework of the National Recovery and Resilience Plan Greece 2.0, funded by the European Union - NextGenerationEU (Implementation body: HFRI) under the research project BEAM-RAISE.}}

\markboth
{Drizari et~al.: Measurement-Driven Learning-Based Beam Selection for Hybrid Beamforming at 26.5 GHz}
{Drizari et~al.: Measurement-Driven Learning-Based Beam Selection for Hybrid Beamforming at 26.5 GHz}

\maketitle
\thispagestyle{empty}

\begin{abstract}
This paper investigates learning-assisted transmit beam selection for indoor millimeter-wave (mmWave) systems operating with hybrid beamforming and joint transmission. A synchronized SDR-based testbed at 26.5 GHz band is deployed to collect wideband channel measurements in a realistic office corridor environment. Using the measurement dataset, beam selection is formulated as a supervised learning problem aiming to approximate the SNR-optimal beam obtained through exhaustive sweeping. Two complementary approaches are examined: a geometry-driven Deep Neural Network (DNN) that predicts the optimal beam from spatial features, and a pilots-only method that infers suitable beams using a limited number of sounded pilot beams without positional information. Experimental results demonstrate high prediction accuracy and significant reduction in beam search overhead compared to exhaustive sweeping, highlighting the effectiveness of measurement-driven learning for practical indoor mmWave beam management.
\end{abstract}



%

\IEEEpeerreviewmaketitle
\section{Introduction}
Millimeter-wave (mmWave) communications are key enablers for future wireless systems, offering large bandwidths and high data rates \cite{Rappaport2017Overview},\cite{Rangan2014Potentials},\cite{Pi2011Introduction}. However, mmWave links are highly dependent on propagation conditions and blockages and are highly sensitive to beam alignment due to increased angular selectivity \cite{10506237}, particularly for rich scattering environments with frequent obstructions \cite{MacCartney2015Indoor},\cite{Rappaport},\cite{Hur}. Thus, efficient beam management is essential to fully exploit the mmWave potential in practical deployments. Hybrid beamforming architectures have emerged as practical solutions that balance performance and hardware complexity by combining digital processing in the baseband with analog beam steering in the radio-frequency (RF) front end \cite{Alkhateeb_Heath},\cite{Heath},\cite{Molisch2017Hybrid},\cite{Sohrabi2016Hybrid}. Towards this direction, he hybrid beamforming was optimized in \cite{10185619} to maximize the sum rate metric. However, the proposed approach was conducted under the assumption of narrow-band systems. Although hybrid beamforming significantly reduces the number of required RF layers, selecting suitable transmit beam configurations remains challenging, especially in dynamic indoor environments where extensive beam sweep introduces substantial overhead \cite{10000956},\cite{Wangetal},\cite{Hur2013Millimeter}. To address this challenge, data driven and learning based approaches have recently gained attention as means to exploit measurement data and infer suitable beam configurations with reduced complexity \cite{alkhateeb_globecom},\cite{Huang_Zappone_Alexand}. These approaches rely on capturing specific propagation characteristics through measurements that are difficult to exploit through real-time optimization.

Many works  have exploited learning schemes for optimizing hybrid beamforming techniques due to the increased complexity which rises. In  \cite{10004962}, the authors proposed a two-stage algorithm to solve a
wideband analog beamforming problem. Deep reinforcement learning was used to
optimize the phase shifts. However, the algorithm in was designed for a single-user with a single RF-chain. Such a system model is
specific and the designed algorithm is difficult to extend
to a generalized framework. To reduce the complexity of beamforming
training, the authors of \cite{9903646} proposed a deep-learning based beam training method, which uses two neural networks for distance and angle domain selection, respectively.  

Triggered by the aforementioned, in this work, a learning-assisted beam selection scheme is presented and developed through indoor mmWave measurement campaigns. An SDR testbed equipped with multi-domain beamforming capabilities is employed to collect wideband channel measurements in a real-world office corridor environment. The measurement setup enables systematic exploration of beam-dependent channel characteristics through structured beam sweeping and spatial sampling, providing a dataset suitable for data driven analysis. Moreover, a system model is formulated that captures the relationship between beam configurations and the resulting channel and {link quality metrics. Using this model, a learning framework is used to infer suitable transmit beam combinations with the aim of enhancing communication performance while reducing beam search overhead. The effectiveness of the proposed approach is evaluated using experimental data derived from  measurements.


\section{Measurement Setup}
The measurement campaign was conducted in an indoor office corridor environment, representative of typical deployment scenarios for future indoor mmWave Radio Units (RUs). The environment consists of enclosed walls, doors, cabinets, and office equipment, which create rich multipath propagation conditions. Due to the geometry of the corridor, strong reflections from planar surfaces are observed \cite{MacCartney2015Indoor}, which constitute a challenging environment.

\begin{figure}
\centering
\includegraphics[width=1\columnwidth]{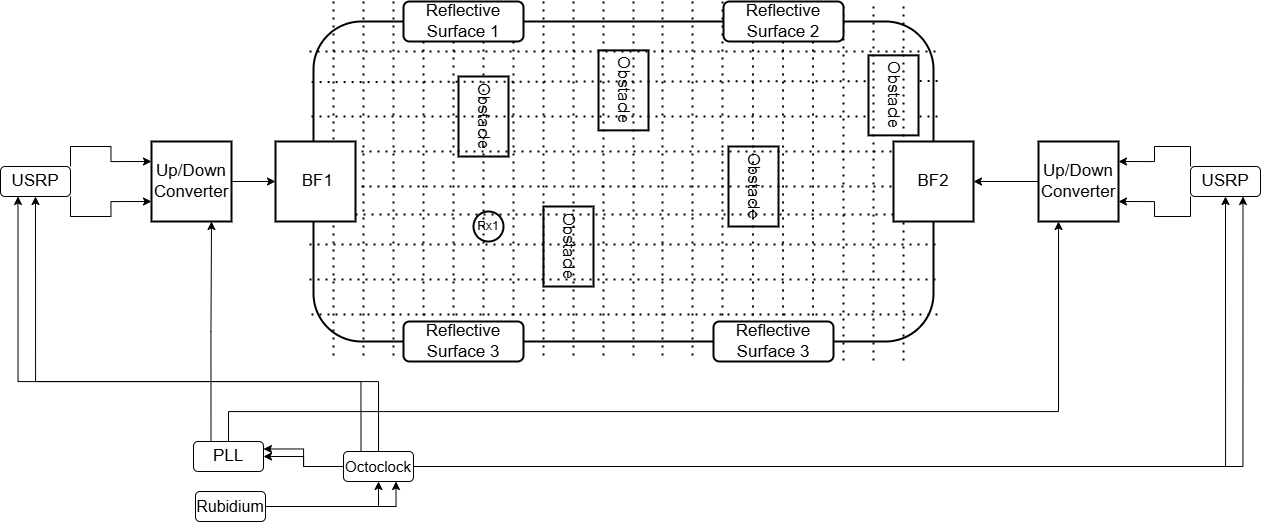}
\caption{Diagram of the measurement campaign setup.}
\label{fig:diagram_setup}
\end{figure}

\begin{figure}
\centering
\includegraphics[width=1\columnwidth]{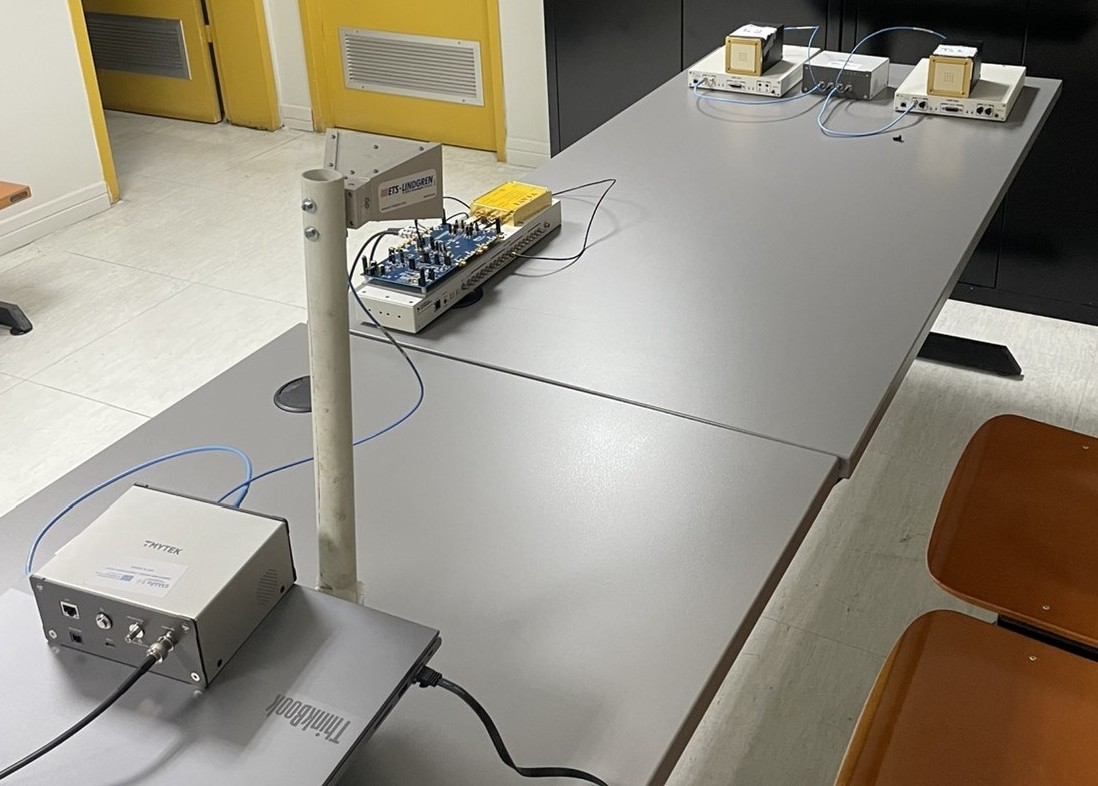}
\caption{View of the measurement equipment components.}
\label{fig:environment}
\end{figure}

\vspace{10pt}
The receiver positions were arranged on a 2D corridor grid to ensure spatially consistent channel sampling. Measurements at these locations were used to train and evaluate an AI-based beam selection algorithm. As receiver position (and orientation) varied across the grid, the learning algorithm inferred the optimal transmit beam pair to enhance communication performance. The system uses two synchronized transmitters (Tx's) with hybrid beamforming capabilities at fixed positions to emulate wall-mounted indoor RUs or base stations. This enables a controlled study of spatial, angular, and beam-dependent channel effects. The resulting dataset supports data-driven beam selection through learning-based optimization. Fig.~\ref{fig:diagram_setup} shows the main components of the measurement setup. The SDR-based hybrid beamforming testbed developed is described in the following.

\subsection{Measurement Equipment and RF Chain}
The testbed consists of two main radio hardware subsystems: (a) the SDR based transmitters, which operate as a multi-RU base station in a cell-free architecture, and (b) the user equipment (UE) mounted on a robotic unit. The main hardware components of the testbed, shown in Fig.~\ref{fig:environment}, are described in the following subsections.

\paragraph{\textbf{Radio Units}}
Each RU includes a dual-channel USRP X310 \cite{Ettus_X310} with UBX-160 daughterboards \cite{Ettus_UBX160}, providing up to 160 MHz bandwidth per channel. Dual-channel mmWave up/down converters (TMYTEK UDBox-5G \cite{TMYTEK_UDBox}) translate the signal to 26.5 GHz, while beamformers (TMYTEK BBox One 5G \cite{TMYTEK_BBox}) enable analog beamforming via phase shifting across 16 RF outputs. To support joint transmission and ensure phase-coherent operation between Tx's, a common reference clock (1 PPS and 10 MHz) is distributed to all RF components using an Ettus OctoClock \cite{Ettus_OctoClock} driven by a high-stability rubidium standard. While USRPs accept 1 PPS and 10 MHz inputs directly, the converters require a 100 MHz reference. A fractional phase-locked loop (PLL) generates this stable 100 MHz signal from the 10 MHz input to drive the converters, ensuring precise time and frequency alignment across the transmit chain.

\subsection{Tx and Rx antennas}

\paragraph{\textbf{Tx Antenna}}
Each Tx is equipped with a commercial 4$\times$4 microstrip patch antenna array \cite{TMYTEK_BBox} provided as part of the TMYTEK antenna array (AA) kit. The AA consists of 16 uniformly spaced radiating elements arranged in a planar configuration operating in the 26 - 29 GHz mmWave band. Each element is individually fed, enabling full control of the amplitude and phase excitation across the array through the beamforming unit. The AA is connected directly to the analog beamformer, which applies programmable gains and phase shifts to the 16 RF paths, allowing electronic beam steering over a predefined angular range. The configuration supports the generation of a multitude of directive beams and facilitates systematic beam sweeping during the measurement campaign. The compact planar structure and directive radiation characteristics of the array make it well suited for indoor RU deployments. The 4$\times$4 AA was selected due to its radiation characteristics, i.e., its high gain (45dB EIRP) and scanning capabilities of $\pm 45^\circ$ in azimuth and elevation, as well as its seamless integration with the hybrid beamforming scheme employed in the testbed. By combining digital precoding with analog beamforming at the RF front end, the Tx antenna system enables controlled exploration of beam dependent channel behavior, essential for evaluating beam selection strategies and learning based optimization methods.

\paragraph{\textbf{Rx Antenna}}
At the receiver, a directive standard gain horn (SGH) antenna is mounted on the mobile UE robotic platform. Operating in the 26.5-28 GHz band, it provides a stable, well-defined high-directivity pattern suitable for controlled measurements. The antenna connects to the RF frontend through the mmWave up/down conversion chain and remains fixed during each measurement. Using a directive, non-beamforming receiver intentionally excludes receive beam adaptation, so variations in signal quality are mainly attributed to transmit beam configuration, Rx orientation, and propagation conditions. The SGH enables repeatable and consistent measurements across the spatial grid. As the robot moves between predefined positions, the antenna samples the channel with a fixed radiation pattern, allowing a reliable comparison of beam-dependent characteristics and supporting the development and evaluation of learning-based transmit beam selection. Additionally, the SGH offers a higher gain ($\approx$ 12 dBi) and eliminates the need for an LNA, simplifying the UE setup and reducing the payload and power requirements on the robotic platform.

\subsection{Baseband operation and Measurement Procedure}
A 160-MHz OFDM waveform, with sampling rate $F_s = 200$ MHz is used for phase coherent indoor channel state information (CSI) measurements under multi beam transmission. The waveform is designed to support and allow for robust synchronization, reliable extraction of complex baseband channel responses, and repeatable beam sweep operation. Each transmission instance consists of a deterministic synchronization preamble followed by an OFDM frame that includes pilot symbols for channel estimation. A short guard interval is inserted between consecutive transmissions to avoid transient effects during beam switching and to ensure clear separation between measurement instances. Reliable synchronization is achieved using a repeated Zadoff--Chu (ZC) sequence that enables correlation-based coarse time synchronization and carrier frequency offset estimation at the receiver. The frame beginning is identified by locating the peak of a normalized correlation metric, while the phase shift between the repeated preamble segments is used to estimate the residual frequency offset. Given the use of shared high stability reference clocks, synchronization between RUs is ensured, while any other offset at the Rx is sufficiently small to be absorbed through equalization. To enable unambiguous beam identification during post processing, beam information is embedded directly into the waveform structure. Each transmit beam is associated with a distinct preamble sequence, enabling the receiver to identify the active beam id by cross-correlating with the preamble set. In the case of two synchronized RUs, each RU contributes independently to the selection of a unique preamble for each joint transmission instance. In the case of two synchronized transmitters, each transmitter independently selects its preamble based on its active beam, resulting in a unique preamble pair for each joint transmission instance.

The measurement environment includes more than 14 blockages and several strong reflectors, including attached passive reflectors (tmytek.com/products/components/xrifle). As a result, beam selection cannot rely on geometric alignment, since the Line-of-Sight path is obstructed or significantly disturbed across most measurement locations.

\section{System Model}
This section presents an extended system model for the conducted hybrid beamforming experiments. The model captures the interaction between digital precoding implemented on the SDRs, and analog beamforming. It is suitable for frequency-selective wide-band systems that form a generic theoretical framework for beam sweeping, beam selection, and rate maximization experiments using multiple transmission points (as RUs in the cell-free context) and a single receiver.

\subsection{OFDM signal transmission}
An OFDM scheme with the following features is considered:
\begin{itemize}[leftmargin=1.5em]
  \item $N$ is the number of OFDM subcarriers and $N_{cp}$ is the length of the cyclic prefix in samples. $F_{S}$ is the sampling frequency and therefore the duration of the OFDM symbol is $T_{OFDM}=\frac{\left(N+N_{c p}\right)}{F_{S}}$, which for $F_S = 200 \mathrm{MHz}, N=4096$ and $N_{c p}=0.125$ results in 0.023 msec. If $M$ OFDM symbols constitute an OFDM frame/slot, scheduling decisions are made per slot - defining the system's Transmission Time Interval (TTI). $N_{null}$ subcarriers are kept at the band edges to reduce bandwidth to 160MHz and reduce sidelobe effects and adjacent channel interference.
 \item Each subcarrier is modulated using QAM with SINR-depended adaptive rank. Link adaptation (i.e., weighting, power control, modulation, and coding) is performed in sets of adjacent subcarriers. Keeping with the 5G terminology, the resource block (RB) contains twelve subcarriers. For each RB, the radio channel is considered approximately flat and pilot subcarriers are allocated for channel estimation and equalization depending on the MIMO scheme.
 \item For an RF layer, the OFDM symbol is given by:
    \begin{linenomath}
    \begin{gather}\label{eq1}
    s(n)=\frac{1}{\sqrt{N}} \sum_{k=0}^{N-1} S(k) e^{2 \pi j \frac{k n}{N}}, 
    \end{gather}
    \end{linenomath}
Then, the symbol is extended with the cyclic prefix to produce $r(n)$.
  \item In a MIMO configuration, depending on the propagation conditions, we can either perform spatial multiplexing or beamforming. In this work, we focus on beamforming in various domains. Network or Distributed MIMO deployments are also considered for cell-free operation. $P$ coordinated simultaneous transmissions are assumed through distributed RUs.
\end{itemize}
In the investigated and measured scenario, a single user signal is ``beamformed'' towards the Rx. Typically, two beamforming modes are considered:

\begin{itemize}
\item Wideband beamforming, where the wideband signal from each source is multiplied by a complex gain to ``shift'' the pattern to specific directions
\begin{linenomath}
\begin{equation}\label{eq2}
\boldsymbol{x}(n)=\boldsymbol{w}(n) r(n),
\end{equation}
\end{linenomath}
where $\boldsymbol{x}(n)$ ($L \times 1$) vector is the Tx signal from the $L$ available RF layers and $\boldsymbol{w}(n)$ contains beamforming gains. This type of beamforming is performed efficiently in the analog domain with a phased array configuration.
\item Precoding, where each RB of the wideband signal is multiplied by a suitable complex gain to optimize a metric/KPI at the receiver. Let $R$ be the number of subcarriers per RB and $Q$ the RBs per OFDM, without loss of generality, we can assume that $N=RQ$. This leads to an alternative expression for the OFDM signal where
\begin{linenomath}
\begin{equation}\label{eq3}
s_{l}(n)=\frac{1}{\sqrt{N}} \sum_{q=0}^{Q-1} \sum_{r=0}^{R-1} W_{l}^{*}(q) S(r+R q) e^{2 \pi j \frac{(r+R q) n}{N}}
\end{equation}
\end{linenomath}
$W_{l}(q)$ is the complex precoding gain per resource block for the $l$-th RF layer.
\item The pilot subcarriers used for channel estimation are generally not multiplied by a beamforming weight. In a more accurate expression, the Tx signal is given by
\begin{linenomath}
\begin{equation}\label{eq5}
\begin{gathered}
  {s_l}(n) = \frac{{\text{1}}}{{\sqrt N }}\sum\limits_{q = 0}^{Q - 1} {\left( {\sum\limits_{r \in \Theta } {W_l^*} (q)S(r + Rq){e^{2\pi j\frac{{(r + Rq)n}}{N}}} + } \right.}  \\ 
  \left. {\sum\limits_{r \in \bar \Theta } S (r + Rq){e^{2\pi j\frac{{(r + Rq)n}}{N}}}} \right) \\ 
\end{gathered}
\end{equation}
\end{linenomath}
where $\Theta$ is the set of data subcarriers and $\bar{\Theta}$ is the complementary set, i.e. the pilots. 
\end{itemize}
Both analog beamforming and precoding are considered. Initially, (\ref{eq3}) is transformed into a matrix relationship. $[\boldsymbol{W}]_{q l}=W_{l}(q)$ is the $Q \times L$ precoding matrix ($q$-th RB and $1$-th RF layer). Each time sample for each of the $L$ layers is given by
\begin{linenomath}
{\small
\begin{equation}\label{eq6}
\begin{aligned}
\hat{\mathbf{s}}(n) =
\Bigg[
\Bigg(
\mathbf{W}^{H}
\underbrace{
\begin{bmatrix}
\mathbf{I}_{R}\\
\vdots\\
\mathbf{I}_{R}
\end{bmatrix}}_{Q\ \text{times}}
\Bigg)
\odot
\underbrace{
\begin{bmatrix}
\mathbf{s} & \cdots & \mathbf{s}
\end{bmatrix}}_{L\ \text{times}}
\Bigg]
\begin{bmatrix}
1\\
e^{j2\pi n/N}\\
\vdots\\
e^{j2\pi (N-1)n/N}
\end{bmatrix}.
\end{aligned}
\end{equation}
}
\end{linenomath}
where $\mathbf{s}$ is the channel vector and operator $\odot$ is the Hadamard product. Hybrid beamforming includes both operations. Let's assume that there are $L_{1}$ Digital-to-Analog converters of an SDR producing $L_{1}$ RF layers. Then, each output is beamformed using analog phase shifters producing $L_{2}$ RF outputs. In total, $L=L_{1} L_{2}$ layers are generated. Modifying (\ref{eq2}) to support $L_{1}$ RF ports as input, if $\boldsymbol{w}_{a}^{(l)}(n)$ is the $L_{2} \times 1$ vector with the analogue beamforming complex gains for the $l$-th ( $l \in 0 . . L_{1}-1$ ) input RF port, and if
\[{{\mathbf{W}}_a}(n) = diag\left( {\left[ {{\mathbf{w}}_a^{(0)}(n),{\mathbf{w}}_a^{(1)}(n)...,{\mathbf{w}}_a^{\left( {{L_1} - 1} \right)}(n)} \right]} \right)\]
then
\begin{linenomath}
\begin{equation}\label{eq7}
\boldsymbol{x}(n)=\boldsymbol{W}_{a}(n) \boldsymbol{r}(n),
\end{equation}
\end{linenomath}
This result includes both the analog and digital beamforming gains (see (\ref{eq6}) and (\ref{eq7})).

\subsection{Triple hybrid beamforming}
Modern antennas can shape radiation patterns beyond the degrees of freedom of conventional phased-array steering. Examples include antennas with predefined azimuth patterns \cite{Kanatas} and Electronically Steerable Parasitic Arrays (ESPAR), which generate diverse beams using simple pin-diode or varactor control \cite{Marantis}. This capability allows extending the beamformer with a third stage: beam pattern selection per Tx element., i.e.,
\[
\tilde{\mathbf{x}}(\varphi, \theta, n)=\left[\begin{array}{c}
b_{0}(\varphi, \theta)\\
\vdots \\
b_{L-1}(\varphi, \theta)
\end{array}\right] \mathbf{x}(n)
\]
where $\widetilde{\boldsymbol{x}}(\varphi, \theta, n)$ represents the Tx signal after leaving the antenna elements in the direction $\varphi$ (azimuth), and $\theta$ (elevation) and $b_{l}(\varphi, \theta)$ is the selected beam for the $l$-th element. It becomes clear that this type of optimization is a cumbersome process. Moreover, the use of pilots to estimate all possible combinations in order to select the best is neither practical nor feasible. However, by continuous monitoring of the radio environment as the antenna switches patterns, training of a learning model becomes possible, selecting a near-optimum solution for a given environment.

\subsection{The Joint Optimization Problem}
The wideband mmWave channel impulse response with multiple propagation paths indexed by $i$, and multiple RUs indexed by $p$, as perceived by a single-RF UE, is written as
\begin{linenomath}
\begin{equation}\label{eq8}
h[m] = \sum\limits_{p = 1}^P {\sum\limits_{i = 1}^J {\left( \begin{gathered}
  {a_{i,p}}{\mathbf{b}}_{{\text{UE}}}^*\left( {\phi _i^{{\text{AoA}}},\theta _i^{{\text{AoA}}}} \right) \cdot  \\ 
  {\mathbf{b}}_{{\text{RU}}}^{(p)}\left( {\phi _i^{{\text{AoD}}},\theta _i^{{\text{AoD}}}} \right)\delta \left( {m - {m_i}} \right){e^{j{\phi _{{D_i}}}}} \\ 
\end{gathered}  \right)} } 
\end{equation}
\end{linenomath}
where $a_{i, r}$ the complex path gain of path $i$ from RU $p$, $m_{i}$ the discrete path delay (in samples), ${{\mathbf{b}}_{{\text{UE}}/{\text{RU}}}}( \cdot ) \in {\mathbb{C}^{{N_{{\text{UE}}/{\text{RU}}}} \times 1}}$ are the array responses and $\phi_{D_{i}}=2 \pi f_{D_{i}} m T_{s}$ the Doppler-induced phase. The block fading assumption is adopted (negligible time variance during an OFDM symbol); thus, the Doppler effect is modeled as a simple phase shift. Applying an $N$-point FFT to $\mathbf{h}[m]$ yields the frequency-domain channel at subcarrier $k$:
\begin{linenomath}
\begin{equation}
H[k] = \sum\limits_{p = 1}^P {\sum\limits_{i = 1}^J {\left( \begin{gathered}
  {a_{i,p}}{{\mathbf{b}}_{{\text{UE}}}^*}\left( {\phi _i^{{\text{AoA}}},\theta _i^{{\text{AoA}}}} \right) \\ 
   \cdot {\text{ }}{\mathbf{b}}_{{\text{RU}}}^{(p)}\left( {\phi _i^{{\text{AoD}}},\theta _i^{{\text{AoD}}}} \right){e^{ - j\frac{{2\pi }}{N}k{m_i}}}{e^{j{\varphi _{\text{D}_i}}}} \\ 
\end{gathered}  \right)} }
\end{equation}
\end{linenomath}
According to the system model, a setup with one or multiple RUs capable of switching patterns is considered. Therefore, each RU employs a beam selected from a predefined codebook. The combined effect of $J$ path gains and delays in the frequency domain is captured by the diagonal matrix, i.e.,
\begin{linenomath}
\begin{equation}
\widetilde{\mathbf{A}_p}(k)=\operatorname{diag}\left(a_{1,p} e^{-j \frac{2 \pi}{N} k m_{1}}, \ldots, a_{J,p} e^{-j \frac{2 \pi}{N} k m_{J}}\right)
\end{equation}
\end{linenomath}
The vector $L \times 1$ ${\mathbf{b}}_{RU}^p\left( {\varphi ,\theta } \right)$ contains the antenna response at angles $\left( {\varphi ,\theta } \right)$ for the $p-$th RU. Similarly, ${\boldsymbol{\beta }}_{RU}^p\left( l \right)$ for $l=1...L$ is defined as the $J \times 1$ vector containing the response of the l-th pattern towards/from all $J$ path angles. Consequently, ${\mathbf{B}}_{RU}^p$ is the matrix that contains the responses for all $J$ paths:

\begin{linenomath}
\begin{equation}
\begin{gathered}
  {\mathbf{B}}_{RU}^p = \left[ {\begin{array}{*{20}{c}}
  {{\mathbf{b}}{{_{RU}^p}^{\text{H}}}\left( {\phi _{1,p}^{{\text{AoD}}},\theta _{1,p}^{{\text{AoD}}}} \right)} \\ 
   \vdots  \\ 
  {{\mathbf{b}}{{_{RU}^p}^{\text{H}}}\left( {\phi _{J,p}^{{\text{AoD}}},\theta _{J,p}^{{\text{AoD}}}} \right)} 
\end{array}} \right]=\left[ {\boldsymbol{\beta} _{RU}^p\left( 1 \right) \ldots \boldsymbol{\beta} _{RU}^p\left( L \right)} \right], \\ 
  {\mathbf{b}}_{RU}^p\left( {\phi _{1,p}^{{\text{AoD}}},\theta _{1,p}^{{\text{AoD}}}} \right) = \left[ {{b_0}\left( {\phi _{1,p}^{{\text{AoD}}},\theta _{1,p}^{{\text{AoD}}}} \right)...{b_{L - 1}}\left( {\phi _{J,p}^{{\text{AoD}}},\theta _{1,p}^{{\text{AoD}}}} \right)} \right]^{H} \\ 
  {\boldsymbol{\beta }}_{RU}^p\left( l \right) = {\left[ {{b_l}\left( {\phi _{1,p}^{{\text{AoD}}},\theta _{1,p}^{{\text{AoD}}}} \right)...{b_l}\left( {\phi _{J,p}^{{\text{AoD}}},\theta _{J,p}^{{\text{AoD}}}} \right)} \right]^H}
\end{gathered}
\end{equation}
\end{linenomath}
we can algebraically express the channel responses for all possible patterns as  ${\mathbf{H}}\left( k \right) = \sum\limits_{p = 1}^P {{\mathbf{b}}_{UE}^H{{{\mathbf{\tilde A}}}_p}\left( k \right){\mathbf{B}}_{RU}^p}$, where ${\mathbf{b}}_{UE}$ is the array respond matrix at the Rx (no beamsteering is considered at the UE). It is assumed that ${{\mathbf{W}}_a}\left( n \right) = {{\mathbf{W}}_a}$ is constant over an OFDM-symbol period. The expressions derived from the system model are general and cover a multitude of triple hybrid beamforming scenarios in terms of RF layers, number of RUs and array designs. In the following, the model is tailored to the measurement campaigns.

\subsection{Mapping the system model to the experimental setup}
For the setup of the experiment, two RUs were available ($P=2$). Each RU has only one RF layer, but it can select one of $L=51$ patterns. Moreover, it is assumed that the RB size is $R=12$ subcarriers, and for $N_{FFT}=4096$, $Q=273$ is the number of RBs, similarly to the 5G waveform. The RUs perform coordinated multi-point transmission towards the UE. According to the previous analysis, the Rx signal at the UE in the frequency domain for the $k-$th subcarrier is 
\begin{linenomath}
\begin{equation}
Y(k) = \sum\limits_{p = 1}^P {{\mathbf{b}}_{UE}^H{{{\mathbf{\tilde A}}}_p}\left( k \right){\boldsymbol{\beta }}_{RU}^p\left( {{l_p}} \right)W_{d,p}\left( {\left\lfloor {\frac{k}{R}} \right\rfloor } \right)X\left( k \right)}    
\end{equation}
\end{linenomath}
where $l_p$ is the selected beam and $W_{d,p}$ is the digital precoding coefficient for the $p$-th RU for the $q=\left\lfloor {\frac{k}{R}} \right\rfloor$ RB. Since $L_1$=$L_2$=1, the analog beamformer $W_\alpha$ is absorbed by the digital precoder. The rate for the wideband channel is given by ($\sigma^2$ is the noise variance);
\begin{linenomath}
\begin{equation}
\begin{gathered}
  {\text{Total Rate}} = \sum\limits_{k = 1}^N {{\text{Rate}}\left( k \right)}  =  \\ 
  \sum\limits_{k = 1}^N {{{\log }_2}\left( {1 + \frac{{{{\left| {\sum\limits_{p = 1}^P {{\mathbf{b}}_{UE}^H{{{\mathbf{\tilde A}}}_p}\left( k \right){\boldsymbol{\beta }}_{RU}^p\left( {{l_p}} \right)W_{d,p}^*\left( {\left\lfloor {\frac{k}{R}} \right\rfloor } \right)} } \right|}^2}}}{{{\sigma ^2}}}} \right)}  \\ 
\end{gathered} 
\end{equation}
\end{linenomath}
The objective is to maximize the total rate.
\begin{linenomath}
\begin{equation}
\begin{gathered}
  \bm{\hat{\theta}} = \mathop {\arg \max }\limits_{\begin{array}{*{20}{c}}
  {\left( {{l_1},{l_2}} \right) \in [1...L]} \\ 
  {{{\mathbf{w}}_{d,1}},{{\mathbf{w}}_{d,2}}} 
\end{array}} \left( {{\text{Total Rate}}\left( {{\bm{\theta}} = \left( {{l_1},{l_2},{{\mathbf{w}}_{d,1}},{{\mathbf{w}}_{d,2}}} \right)} \right)} \right) \hfill \\
  {\text{s}}{\text{.t }}{\mathbf{w}}_{d,p}^H{{\mathbf{w}}_{d,p}}{\text{  =   }}\sum\limits_{q = 1}^Q | {W_{d,p}}(k){|^2} \leq P,{\text{  for all RUs}} \hfill \\ 
\end{gathered}
\end{equation}
\end{linenomath}
where $P$ is used as a power constraint for the digital beamforming task. In order to simplify the solution of the problem, we split the beam selection and precoding processes. \emph{Step 1}: Select the beam combination that maximizes wideband SINR, \emph{Step 2}: Perform Maximal Ratio Transmission (MRT) precoding for the selected beam combination. Therefore, we need to search for the optimal beam combo, while the MRT will define the precoding weights using the pilot estimates.

\section{Learning}
This section presents a learning framework to approximate the optimal selection of the transmit beam without exhaustive sweeping. Based on the measurement scenario in Section II, the problem is formulated as supervised learning, with the aim of predicting the beam configuration that maximizes the received wideband SNR. Two approaches are examined: (i) \emph{a geometry-driven Deep Neural Network (DNN) classifier using spatial and orientation features}, and (ii) \emph{a pilot-based SNR-map regression method} that relies on a reduced set of sounded beams, without requiring positional information. As mentioned, the measurement environment is exceptionally complicated with a plethora of blockages and refelective surfaces, which renders simple geometric beam definition insufficient.

\subsection{DNN-Based Transmit Beam Selection}
The DNN-based transmit beam selection methodology is divided into three main processes: i) dataset generation and feature construction, ii) KPI computation and labeling, and iii) DNN training and online inference~\cite{Tsipi2024}.

\begin{enumerate}
    \item \textbf{Dataset generation and feature construction:}

    A measurement-driven dataset is formed by collecting $N'$ channel snapshots at different receiver locations and orientations. For the $n$-th snapshot, a feature vector $\boldsymbol{\chi}_n$ is constructed to capture the transmitter--receiver geometry and the receiver orientation, i.e., $\boldsymbol{\chi}_n=\big[x_R,\;y_R, \;z_R,\;x_T,\;y_T,\;z_T,\;D,\;\phi_R,\;\theta_R\big]$, where $(x_R, y_R, z_R)$ and $(x_T, y_T, z_T)$ denote the Rx and Tx coordinates, respectively, and $D$ is the three-dimensional Euclidean distance. To improve learning stability, each feature is standardized using the training set statistics $x_i^{\mathrm{norm}}=\frac{x_i-\mu_i}{\sigma_i}$. Moreover, filter-based feature selection (Information Gain and Chi-Squared) was used to obtain a reduced feature representation. Hence, the input vector is defined as
    \begin{linenomath}
    \begin{equation}\label{eq:feat_sel}
    \boldsymbol{\chi}_n=\big[x_R,\;y_R,\;D,\;\phi_R\big],
    \end{equation}
    \end{linenomath}
 which reduces input dimensionality and improves the generalizability of the proposed DNN model~\cite{Tsipi2025}.

    \item \textbf{KPI computation and labeling:}

    Let $\mathcal{B}=\{b_1,\ldots,b_T\}$ denote the Tx-beam codebook. For each feature vector $\boldsymbol{\chi}_n$, the objective of beam selection is to identify the beam index per RU that maximizes the rate, measured via SNR:
    \begin{linenomath}
    \begin{equation}\label{eq:beam_opt}
    b^{\star}(\boldsymbol{\chi}_n)=\arg\max_{b_t\in\mathcal{B}}~\mathrm{SNR}(\boldsymbol{\chi}_n,b_t).
    \end{equation}
    \end{linenomath}
    Consequently, the supervised dataset is
    \begin{linenomath}
    \begin{equation}\label{eq:dataset}
    \mathcal{D}=\{(\boldsymbol{\chi}_n,\; b^{\star}(\boldsymbol{\chi}_n))\}_{n=1}^{N'},
    \end{equation}
    \end{linenomath}
    where the class label corresponds to the optimal beam index.

    \item \textbf{DNN Training and Online Inference: }

    Beam selection is formulated as a multi-class classification problem with $B_T$ candidate transmit beams. To model the nonlinear relationship between the input features and the optimal beam index, a fully-connected Deep Neural Network (DNN) is employed. The hidden layers utilize Rectified Linear Unit (ReLU) activation functions, while the output layer adopts a SoftMax classifier to produce posterior probabilities over all beam classes. Specifically, let $f_{\boldsymbol{\Theta}}(\cdot)$ denote the nonlinear mapping implemented by the trained DNN, parameterized by the set of weights and biases $\boldsymbol{\Theta}$. Given an input feature vector $\boldsymbol{\chi}$, the network outputs a vector of logits $\mathbf{c}=[c_1,\ldots,c_{B_T}]$. The posterior probability of selecting beam $b_t$ is computed as
    \begin{linenomath}
    \begin{equation}\label{eq:softmax}
    \hat{p}_t(\boldsymbol{\chi})=\frac{e^{c_t}}{\sum_{v=1}^{B_T} e^{c_v}}, 
    \qquad t\in\{1,\ldots,B_T\}.
    \end{equation}
    \end{linenomath}
    Accordingly, the predicted transmit beam is obtained via
    \begin{linenomath}
    \begin{equation}\label{eq:predict}
    \hat{b}(\boldsymbol{\chi})=\arg\max_{t}\hat{p}_t(\boldsymbol{\chi}).
    \end{equation}
    \end{linenomath}
    During training, the network parameters $\boldsymbol{\Theta}$ are optimized by minimizing the SoftMax cross-entropy loss using the Adam optimizer. To improve generalization and prevent overfitting, dropout and validation-based early stopping are applied. After offline training, online beam selection is performed through a single forward pass of the normalized input feature vector.
    \begin{linenomath}
    \begin{equation}\label{eq:final_rule}
    \hat{b}(\boldsymbol{\chi})=\arg\max_{t}
    \mathrm{SoftMax}\!\big(f_{\boldsymbol{\Theta}}(\boldsymbol{\chi}^{\mathrm{norm}})\big)_t,
    \end{equation}
    \end{linenomath}
 thus approximating the optimal exhaustive-search solution in \eqref{eq:beam_opt} with significantly reduced computational complexity.
\end{enumerate}

\subsection{Structural Design of the Evaluated Learning Models}
To evaluate learning-based transmit beam selection, several ANN and DNN architectures with increasing depth and representational capacity are investigated. All models are fully connected feed-forward networks (MLPs) trained for multi-class classification, where each class represents an optimal transmit beam from the effective codebook. The input layer matches the feature vector dimension, while the output layer uses 18 SoftMax neurons corresponding to the 18 beam classes in the dataset.
   
    \begin{enumerate}
        \item \textbf{Shallow ANN architectures: }

        The shallow models consist of a single hidden layer and serve as baseline solutions with reduced computational complexity: (i) \textbf{ANN-64:} One hidden layer with 64 neurons and dropout rate 0.10; (ii) \textbf{ANN-128:} One hidden layer with 128 neurons and dropout rate 0.12. These architectures evaluate the ability of low-depth networks to learn the nonlinear mapping between spatial/orientation features and the optimal transmit beam.

        \item \textbf{DNN architectures: }

        To study depth effects, architectures with 2 to 4 hidden layers are evaluated.
            \begin{itemize}[leftmargin=3em]
            \item \textbf{DNN-128-64:} Two hidden layers with 128 and 64 neurons, dropout 0.15.
            \item \textbf{DNN-256-256:} Two hidden layers with 256 and 256 neurons, dropout 0.20.
            \item \textbf{DNN-256-128-64:} Three hidden layers with 256, 128, and 64 neurons, dropout 0.25.
            \item \textbf{DNN-256-256-128-64:} Four hidden layers with 256, 256, 128, and 64 neurons, dropout 0.30.
            \end{itemize}
        All hidden layers employ Rectified Linear Unit (ReLU) activations, while the output layer uses SoftMax to model posterior beam probabilities. Dropout regularization is applied after each hidden layer, with progressively higher dropout rates assigned to deeper architectures to mitigate overfitting.

        \item \textbf{Problem Formulation: }

        Tx's operate with predefined beam codebook $\mathcal{B} = \{ b_1, b_2, \dots, b_{B_T} \},$ where ${B_T}$ denotes the number of candidate transmit beams. For each measurement snapshot $n$, the wideband received SNR is evaluated for all beam configurations in $\mathcal{B}$. The optimal beams in terms of rate per RU are equivalent to SNR optimal beams and defined as
        \begin{linenomath}
        \begin{equation}
        b_n^* = \arg\max_{b \in \mathcal{B}} \mathrm{SNR}_n(b),
        \end{equation}
        \end{linenomath}
        where $\mathrm{SNR}_n(b)$ denotes the measured wideband SNR corresponding to beam $b$ in snapshot $n$. Let the input feature vector be $\boldsymbol{\chi}_n = [x_R, y_R, D, \phi_R]^T \in \mathbb{R}^d$, The supervised dataset is defined again as $\mathcal{D} = \{ (\boldsymbol{\chi}_n, b_n^*) \}_{n=1}^{N'},$ where $N'$ is the total number of measurement instances. The objective is to learn a mapping$ f_{\Theta} : \mathbb{R}^d \rightarrow \mathbb{R}^{B_T}$ that approximates the optimal beam selection rule.
    
    \end{enumerate}

\subsection{Pilots-only Beam Selection via SNR-Map Regression}
This subsection considers a practical regime where explicit geometry is unavailable or undesirable, but limited online sounding is feasible. The goal is to avoid exhaustive sweeping over the full transmit codebook while selecting a near-optimal data beam. We show that sounding only a small set of $\Pi$ pilot beams is sufficient to infer (i) a near-optimal beam per RU and (ii) a beam pair that performs well jointly across both RUs.

\begin{enumerate}
    \item \textbf{Problem formulation and evaluation protocol: }

    Let the transmit codebook contain $B=51$ candidate beams per RU. During dataset construction, we obtain ground-truth per-beam wideband SNR labels from an exhaustive sweep. At inference time, the input consists only of the measured complex channel responses of $\Pi$ pilot beams (evenly spaced in the 51-beam codebook), with $\Pi\leq 7$. We train one regressor per RU to predict the full $B$-dimensional SNR vector, and select the data beam as the argmax of the predicted SNR. More explicitly, for each snapshot and RU $p\in\{1,2\}$, the exhaustive sweep provides a label vector $\mathbf{s}_p\in\mathbb{R}^{B}$ where $[\mathbf{s}_p]_b$ is the wideband SNR associated with beam $b$. At runtime we observe only the complex pilot responses on a subset of beams $\mathcal{P}$, $|\mathcal{P}|=\Pi$, from which we build a real-valued feature vector $\boldsymbol{\chi}_p$. A learned mapping $f_p(\cdot)$ produces an estimate $\widehat{\mathbf{s}}_p=f_p(\boldsymbol{\chi}_p)$, and the selected data beam is
    \begin{linenomath}
    \begin{equation}
    \widehat{b}_p = \arg\max_{b\in\{1,\dots,B\}}\ \widehat{s}_{p,b}.
    \end{equation}
    \end{linenomath}
    Although the two regressors operate independently (one per RU), we evaluate the resulting pair $(\widehat{b}_1,\widehat{b}_2)$ using a strict JOINT metric (defined below) that requires both links to be near-optimal in the same snapshot. To avoid spatial leakage, the train/test split is defined via coarse 2D grid bins (spatial hold-out); receiver coordinates are used \emph{only} to form the split and are never used as model inputs. Performance is measured with a tolerance-based metric: a prediction is successful if the selected beam is within $\Delta$~dB of the optimal-beam SNR. We report per-RU success and a stricter \emph{JOINT} metric that requires both RUs to be within tolerance in the same snapshot, for $\Delta\in\{1,3\}$~dB.
    
     Concretely, if $s^\star_p=\max_b s_{p,b}$ is the optimal SNR for RU $p$ and $s_{p,\widehat{b}_p}$ is the SNR achieved by the predicted beam, then the within-$\Delta$~dB criterion is $s^\star_p - s_{p,\widehat{b}_p} \leq \Delta$. The JOINT success rate is the fraction of snapshots for which this condition is maintained simultaneously for both RUs. This metric is intentionally stricter than per-RU accuracy, since in joint both links must be sufficiently strong at the same time.

    \item \textbf{Feature design and regression model: }

    The pilots-only feature vector is constructed from the complex pilot beam responses using phase-robust transforms that retain information correlated with beam-dependent SNR while mitigating sensitivity to unknown common phase rotations (e.g., residual CFO / phase noise). For each pilot beam, we extract: (i) normalized magnitudes to capture frequency selectivity and shadowing, (ii) differential phase across adjacent subcarriers to remove common phase while preserving multipath structure, and (iii) CIR magnitude from the delay domain, highlighting dominant taps and their spread. These components are concatenated across the $\Pi$ pilot beams to form the final real-valued feature vector. Per RU, we employ a lightweight linear pipeline (StandardScaler $\rightarrow$ PCA $\rightarrow$ Ridge regression with cross-validated regularization). PCA reduces dimensionality and improves conditioning, and ridge regularization stabilizes the per-beam SNR prediction under finite training data. This design keeps inference fast (a single feature extraction + matrix operations) while still learning a global mapping from pilots to the full $B$-beam SNR landscape. We intentionally predict an \emph{SNR map} (a length-$B$ vector) rather than directly predicting a single beam index. This provides richer supervision during training, supports tolerance-based evaluation (within-$\Delta$~dB), and naturally extends to top-$K$ recommendations for fast refinement with minimal additional probing.

    \item \textbf{Results and overhead reduction: }
    
    The proposed SNR-map regression achieves high JOINT reliability while reducing online sounding from an exhaustive $B=51$-beam sweep to only $\Pi$ pilot-beam measurements. This corresponds to an overhead fraction of $\Pi/51$ (e.g., $\Pi=7$ is 13.7\% of a full sweep). Table~\ref{tab:ml_pilots_sweep} quantifies the reliability--overhead trade-off. As $\Pi$ increases, the model receives a richer ``fingerprint'' of the channel/geometry through the pilots and can more accurately reconstruct the relative SNR ordering across the entire codebook. Two trends are particularly important. First, JOINT performance improves rapidly with a small number of pilots: with only $\Pi=4$ pilots, JOINT within-$3$~dB already reaches 0.833, indicating that a small pilot set can reliably identify a near-optimal beam pair for joint. Second, the stricter within-$1$~dB metric shows the expected behavior: it is lower for small $\Pi$, but rises sharply once enough pilots are available to resolve finer differences between beams (e.g., 0.685 at $\Pi=5$ and 0.826 at $\Pi=6$). With $\Pi=7$ pilots, the method attains 0.927 JOINT within-$3$~dB and 0.831 JOINT within-$1$~dB under spatial hold-out, with diminishing returns beyond $\Pi\approx 5$--$6$.

        \begin{table}
        \caption{Pilots-only SNR-regression beam selection: JOINT success rate versus the number of pilot beams (spatial hold-out, K=1).\label{tab:ml_pilots_sweep}}
        \centering
        \footnotesize
        \setlength{\tabcolsep}{4pt}
        \renewcommand{\arraystretch}{1.05}
        \begin{tabular}{c r r r r}
        \toprule
        \multicolumn{1}{c}{$\Pi$ pilots} & \multicolumn{1}{c}{JOINT ($\Delta=3$ dB)} & \multicolumn{1}{c}{Inc.} & \multicolumn{1}{c}{JOINT ($\Delta=1$ dB)} & \multicolumn{1}{c}{Inc.}\\
        \midrule
        1 & 0.354 & -- & 0.159 & --\\
        2 & 0.562 & +0.208 & 0.279 & +0.120\\
        3 & 0.638 & +0.076 & 0.367 & +0.088\\
        4 & 0.833 & +0.195 & 0.523 & +0.156\\
        5 & 0.885 & +0.052 & 0.685 & +0.162\\
        6 & 0.914 & +0.029 & 0.826 & +0.141\\
        7 & 0.927 & +0.013 & 0.831 & +0.005\\
        \bottomrule
        \end{tabular}
        \end{table}

    \item \textbf{Discussion and next steps: }

        These results indicate that a small number of pilot beams contains enough information to reconstruct the wideband SNR landscape over the full codebook and to select beams that remain reliable under spatial hold-out. To further strengthen the pilots-only contribution, the next steps are to: (i) evaluate robustness under controlled phase/CFO perturbations and across measurement days (domain shift), (ii) report top-$K$ beam selection (e.g., $K\in\{2,3\}$) to support fast refinement with minimal additional probing, and (iii) evaluate a pilots-only DNN using the same spatial split and within-$\Delta$~dB / JOINT metrics.

\end{enumerate}

\section{Results}

\subsection{Link quality metrics}
This subsection presents a statistical characterization of the measured indoor channel using extracted SNR values at all receiver locations, antenna orientations, and transmit beam combinations. The analysis relies on empirical distributions from recorded I/Q samples and provides insight into the variability and reliability of the mmWave link.

The curve in Fig.~\ref{fig:Coverage_prob} shows the fraction of measurements that exceed a target SNR. As the threshold increases, the coverage decreases monotonically, illustrating the trade-off between achievable SNR and spatial coverage. At low thresholds, coverage approaches unity, indicating basic connectivity across most locations and beam combinations. As the threshold increases, coverage decreases since only a subset of spatial positions and beam configurations supports high-SNR operation. This behavior is typical of directional mmWave systems in rich indoor environments, where performance depends strongly on beam alignment and geometry. In addition to the SNR analysis, Fig.~\ref{fig:Constellation} shows the received 16-QAM constellation measured during the campaign. Monitored in real time, it served as a practical validation of the signal chain, including synchronization, frequency offset correction, and channel equalization. The received symbols form well-defined clusters around the ideal constellation points, indicating accurate timing and effective phase and amplitude correction. The limited dispersion reflects stable link conditions and sufficient SNR for the selected beams. Overall, the constellation confirms the proper operation of the hybrid beamforming testbed and the integrity of the baseband data used for SNR extraction and learning-based beam selection.

\begin{figure}
    \centering
    \includegraphics[width=0.95\columnwidth]{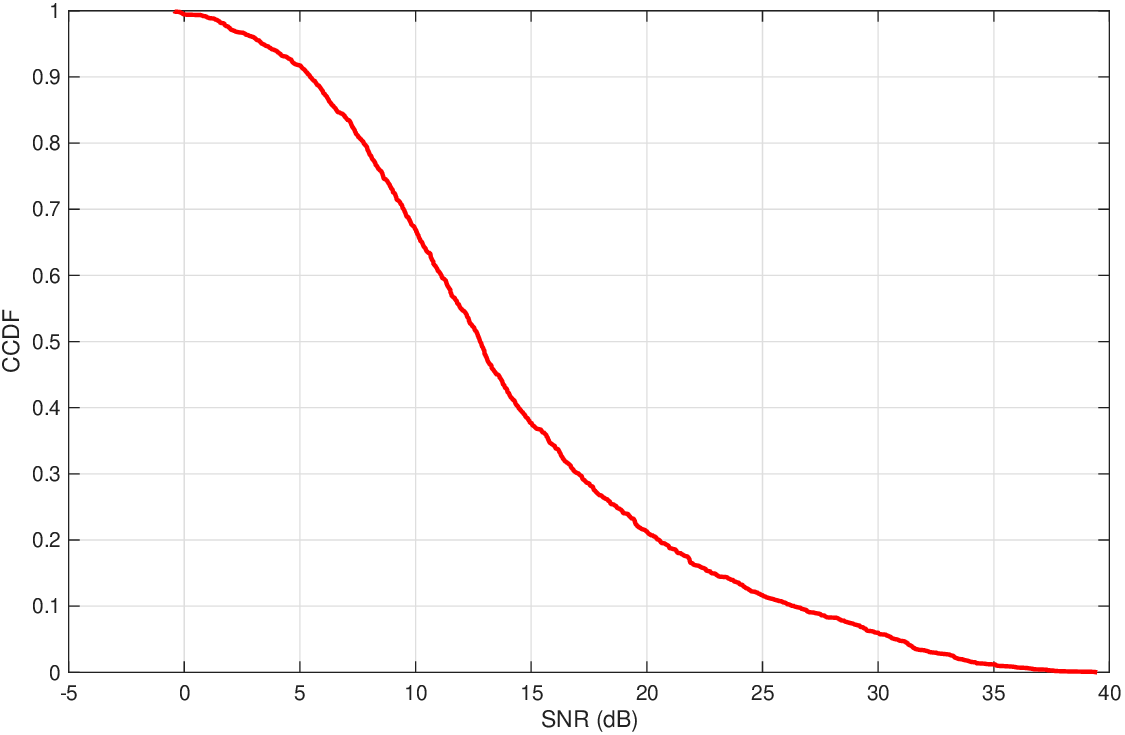}
    \caption{SNR based coverage probability from measurement data.}
    \label{fig:Coverage_prob}
\end{figure}

\begin{figure}
    \centering
    \includegraphics[width=0.9\columnwidth]{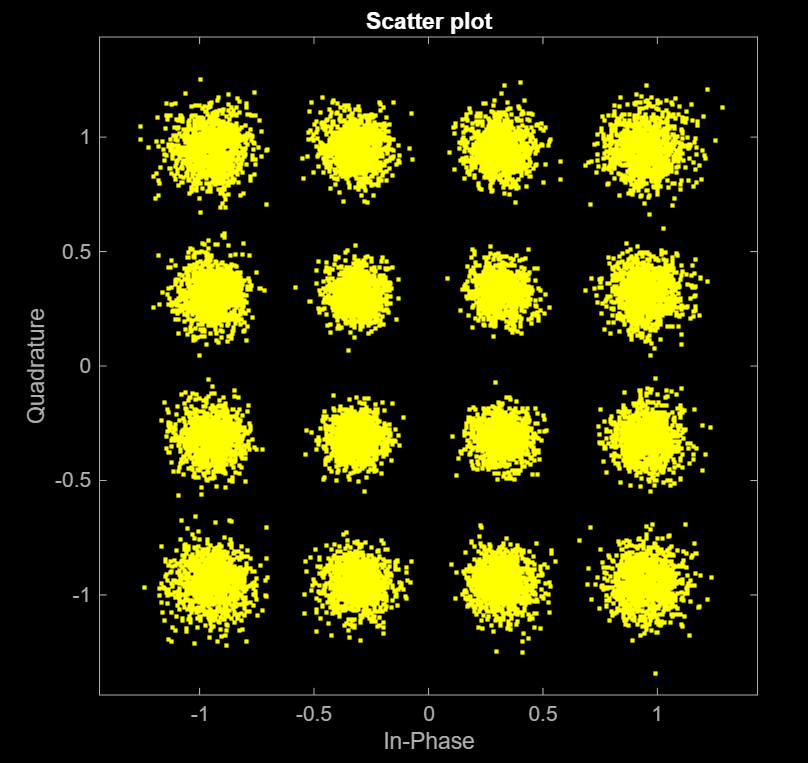}
    \caption{Received constellation of the 16 QAM transmitted data.}
    \label{fig:Constellation}
\end{figure}

\subsection{Geometry driven DNN}
The performance of learning-based beam selection models is evaluated on a held-out test set to assess generalization to unseen receiver positions and antenna orientations. Since the task is formulated as multi-class classification, evaluation metrics include accuracy, precision, recall, and F1-score. Fig.~\ref{fig:test_accuracy} shows that all ANN/DNN models achieve high accuracy ($>93\%$), confirming the effectiveness of the spatial and orientation-aware features. Among shallow models, ANN-64 attains a precision of 93.30\%, increasing to 93.70\% for ANN-128. Introducing moderate depth (DNN-128-64) yields only marginal gains. The best performance is achieved by DNN-256-256 with 94.50\% accuracy, demonstrating the advantage of higher representational capacity in modeling nonlinear feature--beam relationships. Deeper models (DNN-256-128-64 and DNN-256-256-128-64) show saturation without further improvement. Fig.~\ref{fig:metrics} reports macro-averaged precision, recall, and F1-score. The precision ranges from 92.4--93.6\%, to recall from 92.7--93.9\%, with similar F1 trends. DNN-256-256 provides the best balanced overall performance.

\begin{figure}
\centering
\includegraphics[width=1\columnwidth]{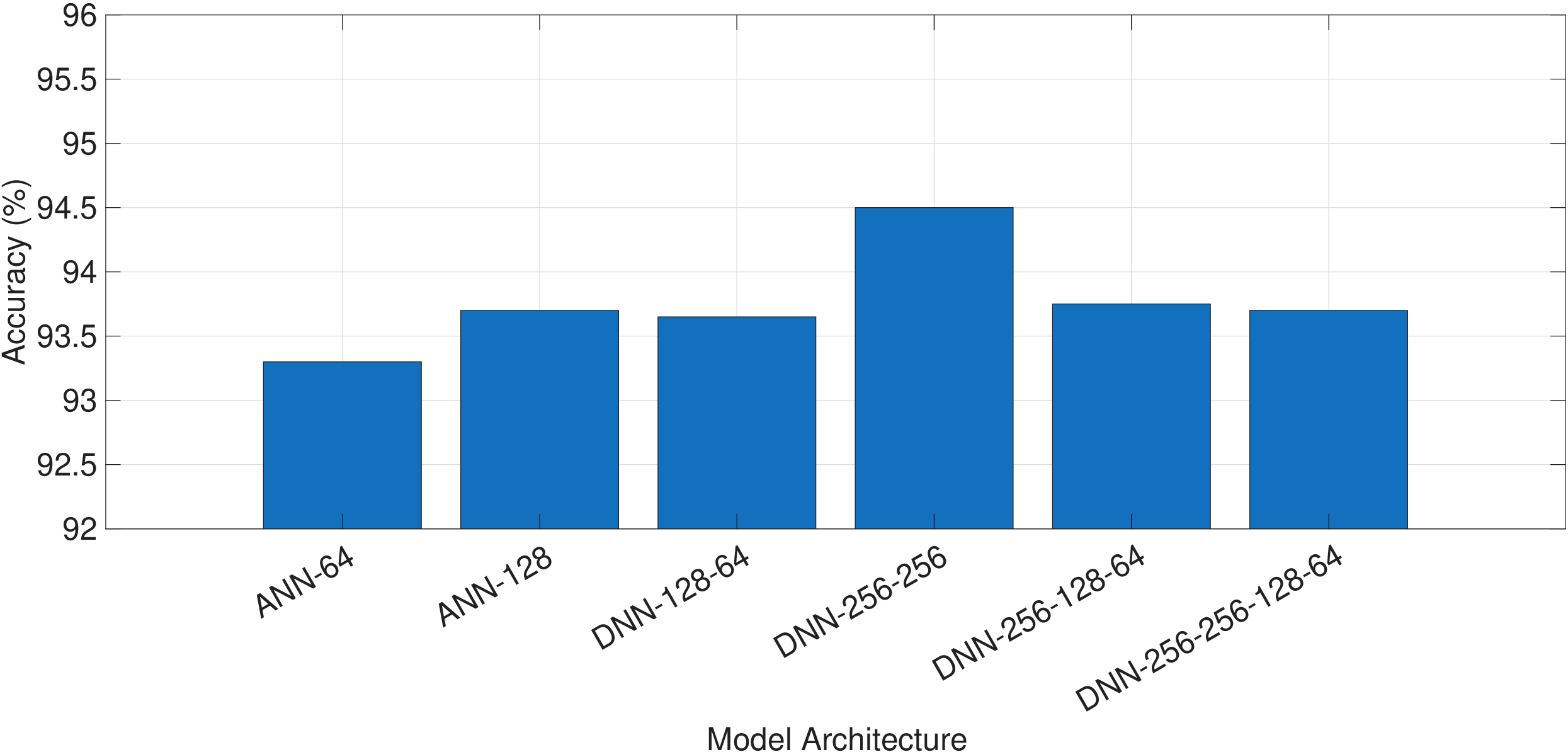}
\caption{Test accuracy achieved by the examined ANN and DNN architectures.}
\label{fig:test_accuracy}
\end{figure}

\begin{figure}
\centering
\includegraphics[width=1\columnwidth]{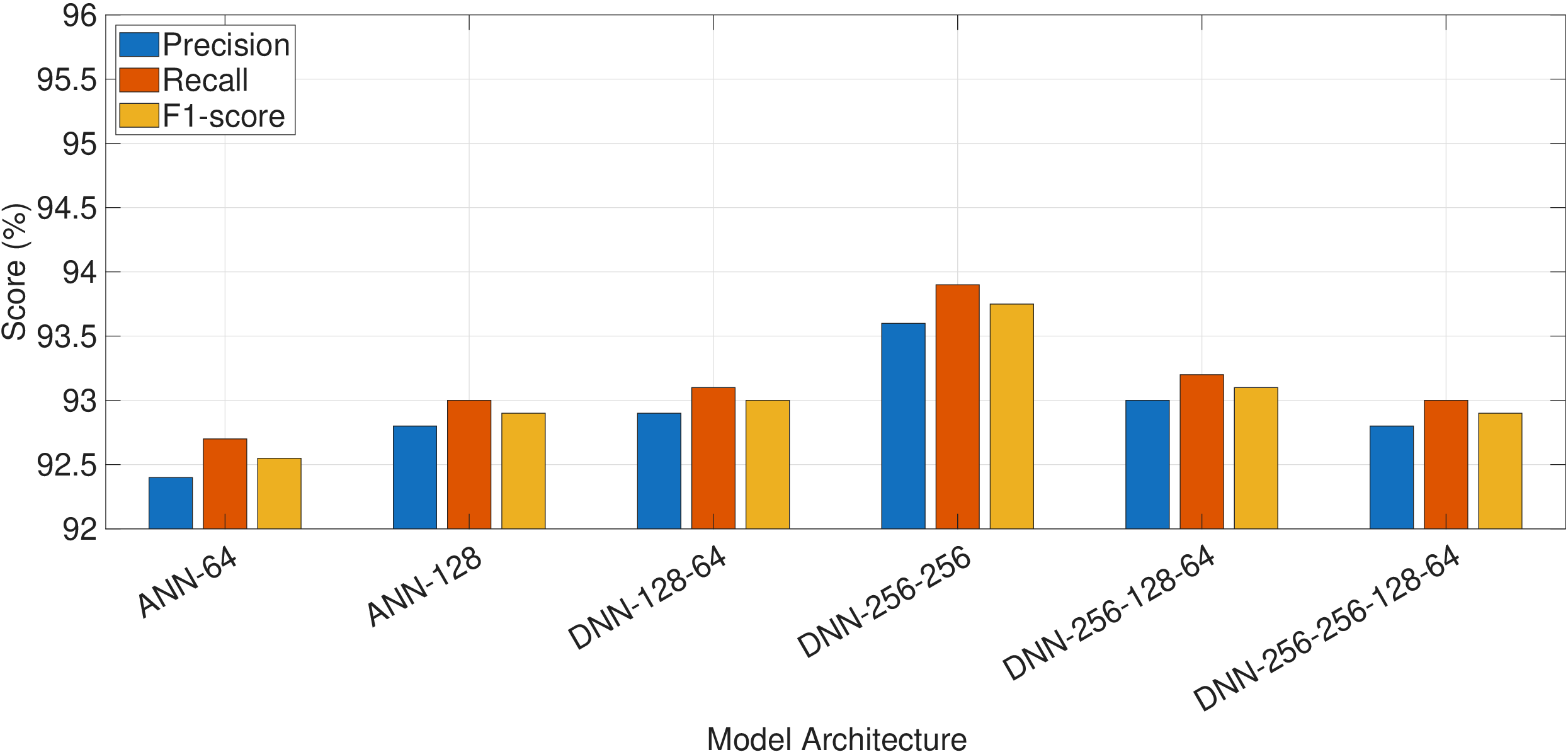}
\caption{Precision, recall, and F1-score comparison on the test dataset.}
\label{fig:metrics}
\end{figure}

\subsection{Pilots-only SNR Regression}
We now summarize the pilots-only beam selection results of the proposed SNR-map regression approach under the spatial hold-out protocol described earlier. In this setting, the transmitter sounds only $\Pi$ pilot beams (evenly spaced in the 51-beam codebook), extracts phase-robust features from the complex pilot responses, and predicts the full $B=51$-dimensional per-beam SNR map per RU. The selected data beam for each RU is the argmax of its predicted SNR map (i.e., $K=1$ selection).

Fig.~\ref{fig:ml_joint_vs_pilots} shows the JOINT within-$3$~dB success rate as a function of the number of beam pilots $\Pi$. The curve increases monotonically, demonstrating that additional pilots translate directly into improved reliability of the inferred beam pair. Importantly, the regime $\Pi=4$--$5$ already yields strong JOINT reliability while using only 7.8\%--9.8\% of the measurements required by a full 51-beam sweep. Table~\ref{tab:ml_pilots_sweep} reports both within-$3$~dB and the stricter within-$1$~dB JOINT metrics: For $\Pi=7$ (13.7\% overhead), the method achieves 0.927 JOINT within-$3$~dB and 0.831 JOINT within-$1$~dB.

\begin{figure}
  \centering
  \includegraphics[width=1\columnwidth]{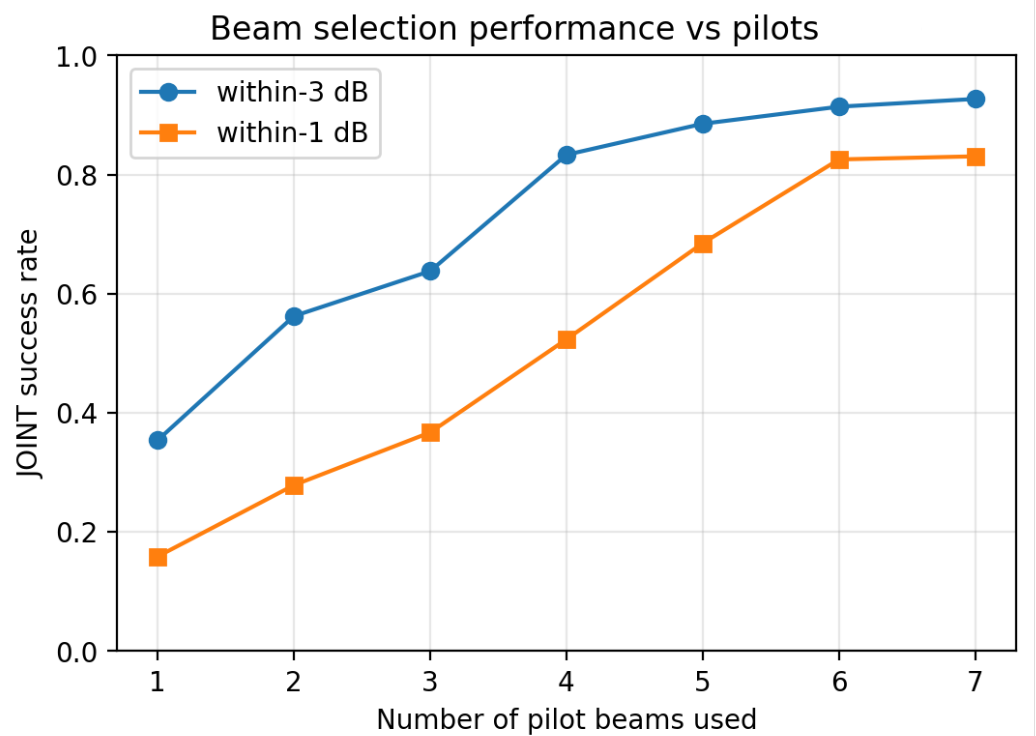}
  \caption{Pilots-only SNR regression: JOINT within-$3$ dB success rate vs. pilot beam number, evaluated under spatial hold-out.}
  \label{fig:ml_joint_vs_pilots}
\end{figure}

\section{Discussion}
The presented results demonstrate that measurement-driven learning can effectively reduce the beam-management overhead of indoor mmWave hybrid beamforming systems. The geometry-driven DNN approach achieves high classification accuracy using spatial and orientation-related input features, showing that receiver location and orientation provide strong information about the most suitable transmit beam in the measured office-corridor environment. At the same time, the pilots-only SNR-map regression approach demonstrates that explicit positional information is not always required. By sounding only a small subset of pilot beams, the method reconstructs the full SNR landscape with sufficient accuracy to identify near-optimal beam pairs under a strict JOINT reliability criterion.

The comparison of the two learning modes highlights complementary operational regimes. Geometry-driven learning is attractive when location or tracking information is available, as it enables beam selection with minimal RF sounding. In contrast, the pilots-only method is more suitable when position information is unavailable, unreliable, or undesirable, while still allowing a substantial reduction in exhaustive beam sweeping overhead. These findings indicate that measurement-based learning frameworks can support practical indoor mmWave beam management, especially in environments with strong angular selectivity, blockages, and reflective surfaces.

\section{Conclusions}
This paper presented a measurement-driven study of learning-assisted transmit beam selection for indoor mmWave systems with hybrid beamforming and joint transmission. An SDR platform at 26.5 GHz collected wideband channel measurements in a realistic office corridor, enabling systematic analysis of beam-dependent propagation. Beam selection was formulated as a supervised learning task to approximate the SNR-optimal beam from exhaustive sweep. Two approaches were evaluated: a geometry-based DNN classifier and a pilots-only method using limited RF sounding. Results show that geometry-aware learning achieves high accuracy at unseen locations, while the pilots-only approach offers a strong trade-off between performance and sounding overhead. In general, the study demonstrates the practical feasibility of data-driven beam management for indoor mmWave and highlights the importance of measurement-based validation for learning-assisted wireless techniques.


\section*{Acknowledgment}
This work was supported in part by the University of Piraeus Research Centre, the EU SNS JU HORIZON Programme through iSEE-6G under Grant 101139291 and within the framework of the National Recovery and Resilience Plan Greece 2.0, funded by the European Union - NextGenerationEU (Implementation body: HFRI) under the research project BEAM-RAISE.

\bibliographystyle{IEEEtran}
\bibliography{references}


\end{document}